\documentstyle[12pt]{article}

\begin{document}

\title{The Problem of $\alpha_s$ in Supersymmetric Unified Theories
\footnote{Supported in part by Department of Energy grant
\#DE-FG02-91ER406267}}

\author{{\bf S.M. Barr}\\ 
Bartol Research Institute\\ University of Delaware\\
Newark, DE 19716}

\date{BA-98-24}
\maketitle

\begin{abstract}

It is shown that in $SO(10)$ there is a general connection
between the suppression of higgsino mediated proton decay
and the value of $\alpha_s$. However, agreement with the experimental
value of $\alpha_s$ can be obtained if there are relatively
large negative threshold corrections to $\alpha_s$ coming from
superheavy split multiplets. It is shown that such split multiplets
can arise in $SO(10)$ without fine-tuning of parameters.

\end{abstract}

\section{Introduction}

One of the strongest pieces of evidence for grand unification is
the fact that the gauge couplings of the minimal supersymmetric
standard model (MSSM) come very nearly together when 
extrapolated to high energy using the renormalization group 
equations [1]. If one were to use the best experimental values for 
$\alpha_s$ and $\alpha$, the experimental value of $\sin^2 \theta_W$ 
and the value predicted 
by unification and the would be discrepant by less than a percent.  
However, it now customary to use the measured values of $\alpha$
and $\sin^2 \theta_W$, since they are the better known quantities, to 
``predict" the value of $\alpha_s$. When expressed in terms of
$\alpha_s$, the agreement with experiment appears somewhat less
dramatic. Indeed, as is well known, there is a bit of a
problem.

The prediction of (minimal) supersymmetric grand unified theories 
(SUSY GUTs) is that $\alpha_s = 0.127 \pm 0.005 \pm 0.002$, 
according to the analysis of [2], where 
the first error is the uncertainty in the low-energy sparticle spectrum, 
and the second is the uncertainty in the top quark and Higgs boson
masses. In [3] a more exact treatment of the low-energy thresholds
gave slightly larger values of $\alpha_s$. In particular, it was
found that $\alpha_s (M_Z) > 0.126$ for no unification thresholds,
$m_{\tilde{q}} \leq 1$ TeV, and $m_t = 170$ GeV. 
On the other hand, a global fit of low-energy data gives the 
result: $\alpha_s = 0.117 \pm 0.005$ [4]. One way these numbers can be
brought into better agreement is by threshold corrections at the GUT
scale. Let us define the threshold correction in $\alpha_s$ at the
GUT scale to be $\epsilon_3 \equiv [\alpha_3(M_G) - \tilde{\alpha}_G]/
\tilde{\alpha}_G$, where $M_G$ is the scale at which the gauge
couplings $\alpha_1$ and $\alpha_2$ meet [5]. Then, to bring the 
SUSY GUT prediction of the strong coupling into agreement with
experiment by high-energy threshold corrections requires the
existence of superheavy fields beyond those of minimal $SU(5)$
that contribute $\delta \epsilon_3 \sim -0.02$ to $-0.03$ [5].

Actually, the problem is worse than this because
of the need to suppress the Higgsino-mediated proton decay amplitudes
[6].
The mechanisms which do this naturally tend to give a {\it positive} 
contribution to $\epsilon_3$, as was noted in [7] and [5]. 
Typically, for $\tan \beta \sim 1$ this contribution is about 
$+ 0.02$, and for $\tan \beta \sim 60$ it is about
$+ 0.05$. Thus, for large $\tan \beta$, the other threshold
corrections would have to be about $-0.08$ to
bring $\alpha_s$ into agreement.  

There are two problems with achieving such large negative threshold
corrections. First, the multiplets which are naturally highly split
increase rather than decrease $\alpha_s$. Second, it is difficult to get
the threshold corrections at the GUT scale to be as large as needed
without either having a fine-tuning of the splittings in multiplets,
or having large numbers of multiplets that happen all to contribute
in the same direction. 

Other solutions to the problem have been suggested besides 
threshold effects at the GUT scale [8]. These include corrections
from sparticle thresholds [9], and intermediate-scale 
gauge symmetry breaking [10].
In this paper, we shall look at the problem in the context of
$SO(10)$ with no intermediate gauge symmetry breaking, and we
shall assume that high energy threshold effects account
for the discrepancy. In section 2, we shall review the problem
in more detail. In particular we shall look at the simplest
mechanism for suppressing Higgsino-mediated proton decay and show
why it exacerbates the problem. We shall also show why it
is hard to get threshold effects of the needed sign and 
magnitude without large numbers of multiplets or fine-tuning.

In section 3, we shall prove a theorem that shows that there is
a close connection between the $\alpha_s$ problem and the 
proton-decay problem in a wide class of Higgs sectors. This
theorem will allow us to see what is necessary to solve the
problem without fine-tuning. In particular, we shall see why
it is easy to solve the problem in the context of $SO(10)$
models with two or more adjoint Higgs fields. 

In section 4, we shall examine the question of whether the 
needed threshold corrections can be obtained in a natural
way with only a single adjoint Higgs. Two ways of doing this
will be proposed, one where the VEV of the adjoint is
proportional to $B-L$, and one where it is proportional to
the third component of $SU(2)_R$. These are the two possibilities
which allow a natural doublet-triplet splitting in $SO(10)$.

 \section{The magnitude of the problem}

In $SO(10)$ the only natural way to solve the doublet-triplet
splitting problem appears to be the Dimopoulos-Wilczek
mechanism [11],[7], also known as the missing-VEV mechanism. 
The heart of the mechanism is an adjoint Higgs multiplet
(a ${\bf 45}$), which gets a vacuum expectation value (VEV)
proportional to the generator $B-L$. That is, calling the adjoint $A$,
one has in an obvious notation, 

\begin{equation}
\langle A \rangle = \left( 
\begin{array}{ccccc} 0 & & & & \\ & 0 & & & \\ & & a & & \\
& & & a & \\ & & & & a \end{array} \right) \times i \tau_2,
\end{equation}

\noindent
where $a \simeq M_G.$ 
Suppose that this adjoint couples to a pair of ${\bf 10}$'s, denoted
$T_1$ and $T_2$, as follows,

\begin{equation}
W_{2/3} = g T_1 A T_2 + M (T_2)^2.
\end{equation}

\noindent
The $T_i$ contain $({\bf 1}, {\bf 2}, \frac{1}{2}) + H.c.$
representations, which we denote $d_i + \overline{d}_i$, and
$({\bf 3}, {\bf 1}, -\frac{1}{3}) + H.c.$ representations, which we
denote $t_i + \overline{t}_i$. These have the following
mass matrices due to the terms in Eq. (2):

\begin{equation}
W_2 = (\overline{d}_1, \overline{d}_2) \left( 
\begin{array}{cc} 0 & 0 \\ 0 & M \end{array} \right) \; \left( 
\begin{array}{c} d_1 \\ d_2 \end{array} \right),
\end{equation}

\noindent
and 

\begin{equation}
W_3 = (\overline{t}_1, \overline{t}_2) \left( 
\begin{array}{cc} 0 & a \\ -a & M \end{array} \right) \; \left( 
\begin{array}{c} t_1 \\ t_2 \end{array} \right).
\end{equation}

\noindent
The light doublets, which are the Higgs of the MSSM, are then
just the $\overline{d}_1 + d_1$. (Eq. (3) shows these to be
massless, but they will gets weak-scale mass from other effects
that are not included in the terms of Eq. (2). However, we 
will ignore weak-scale effects.) Proton decay can happen
through dimension-five operators coming from the exchange 
of their color-triplet partners, $\overline{t}_1 + t_1$. The
relevant propagator, at low momenta, is just given by
$(M_3^{-1})_{11}$, where $M_3$ is the two-by-two mass matrix in
Eq. (4). This is just given by $(M_3^{-1})_{11} = (M/a^2)$. 
Since, in the minimal SUSY $SU(5)$ model, the analogous color-triplet
propagator is just given by $1/a \simeq 1/M_G$, the suppression factor 
of the proton-decay amplitude relative to the minimal $SU(5)$ model,
which we shall call $F^{-1}$, is $F^{-1} = M_G (M_3^{-1})_{11}
= M/a$. For $\tan \beta \sim 1$ typically $F^{-1}$ must be about
$1/10$,
while for $\tan \beta \sim 60$ it needs to be of order
$1/600$ [6].

In [5] it is shown that the effect of superheavy matter multiplets on
$\epsilon_3$ is given by

\begin{equation}  
\epsilon_3 = \frac{\tilde{\alpha}_G}{2 \pi} \sum_{\gamma}
\left[ (b_3^{\gamma} - b_2^{\gamma}) - \frac{1}{2} 
(b_2^{\gamma} - b_1^{\gamma}) \right] \ln ( {\rm det'}
(M_{\gamma}/M_G)),
\end{equation}

\noindent
where the index $\gamma$ stands for a specific
type of $SU(3) \times SU(2) \times U(1)$ multiplet, and
$M_{\gamma}$ is the mass matrix of the multiplets of that type.
$\det' M_{\gamma}$ is the determinant of $M_{\gamma}$, unless
$\gamma = ({\bf 1}, {\bf 2}, \frac{1}{2}) + H.c.$, in which case
it stands for that determinant with the weak-scale eigenvalue removed.  
$b_N^{\gamma}$
is the index of the representation $\gamma$ under $SU(N)$.
(That is, $b_N^{\gamma} \equiv {\rm Tr}_{\gamma} (\lambda_N)^2$, where 
$\lambda_N$ is a generator of $SU(N)$ normalized so that $b_N = 1/2$ for
the fundamental representation.)
Let us see what the effect of the vector multiplets $T_1$ and $T_2$
are. From Eq. (4), the determinant of the color-triplet multiplets $t_i + 
\overline{t}_i$ is just $a^2$; while, from Eq. (3), the determinant
(i.e. $\det'$) of the superheavy weak doublets is just $M$. Thus, 
Eq. (5) gives for the contribution to $\epsilon_3$ from the doublet-triplet
splitting sector

\begin{equation}
\delta \epsilon_3 |_{2/3} = \frac{3 \tilde{\alpha}_G}{5 \pi} \; \ln 
(a^2/ M M_G)  = \frac{3 \tilde{\alpha}_G}{5 \pi} \; \ln F.
\end{equation}

\noindent
Notice that the greater the suppression of proton decay --- that is,
the larger F --- the larger is $\alpha_s$. Every increase in $F$ by
a factor of $10$ increases $\epsilon_3$ by about $0.02$. In 
minimal SUSY $SU(5)$, one needs $\epsilon_3$ to be about $-0.02$ to
$-0.03$. Thus, if $F = 600$, the matter other than the 
$T_i$ contribute about $- 0.08$ to $\epsilon_3$.

We shall show in the next section that this relation between 
$\epsilon_3$ and $F$ is fairly general.

The question arises whether there are small sets of multiplets that 
would give a negative contribution to $\epsilon_3$ as large as is
needed. The obvious choice would be additional ${\bf 10}$'s of $SO(10)$.
For ${\bf 10}$'s, as already noted, $\delta \epsilon_3 
|_{10} = (3 \tilde{\alpha}_G/5 \pi) \ln (\det M_3/\det M_2)$, 
where $M_3$ and $M_2$ are the mass matrices of the triplets
and doublets in the ${\bf 10}$'s respectively. Let the mass of the
additional
${\bf 10}$'s come from both explicit mass terms and from the VEV of the 
adjoint $A$. If, for example, there are in addition to 
$T_1$ and $T_2$, a pair $T$ and $T'$, with mass terms $m T^2 +
m' T^{'2} + T A T'$, then the determinant of the mass matrix
of the doublets in $T$ and $T'$ is just $m m'$, while that of
the triplets is $m m' - a^2$. And therefore the logarithm in 
$\delta \epsilon_3|_{T,T'}$ is $\ln (\det M_3/\det M_2) = 
\ln |1 - (a^2/m m')|$. If $a^2 \gg m m'$, the logarithm is large, but
gives a {\it positive} contribution to $\epsilon_3$. On the other hand,
if $m m' \gg a^2$, the logarithm is negative but small. The largest
{\it negative} logarithm one would expect to get, without a fine-tuned
cancellation between $m m'$ and $a^2$, is of order one. Therefore, there
would have to be of order ten pairs of ${\bf 10}$'s, all
happening to contribute negatively, to get the needed $\delta \epsilon_3
\sim -0.08$. This itself is equivalent to a fine-tuning.

One can look at other representations besides vectors, but the 
situation is similar. There has to be either fine-tuning or a large
number of multiplets which all happen to contribute negatively
to $\epsilon_3$. For example, if there is a spinor pair that gets
mass from a term $\overline{{\bf 16}} \langle {\bf 45} \rangle
{\bf 16}$ with $\langle {\bf 45} \rangle \propto B-L$, the
contribution to $\epsilon_3$ is only $- (3 \tilde{\alpha}_G/10 \pi) 
\ln 3 \cong - 4 \times 10^{-3}$.

\section{Relation of $\alpha_s$ to p-decay in general}

The relationship between $\epsilon_3$ and the suppression of
Higgsino-mediated proton decay that exists in the simplest
situation just analyzed is actually quite general. We shall
first illustrate this with another example, and then prove a
general statement.

Consider a model where the doublet-triplet splitting comes from
the following superpotential terms: 
$W_{2/3} = m {\bf 10}_1 {\bf 10}_2 + m' {\bf 10}_2 {\bf 10}_3
+ {\bf 10}_3 {\bf 10}_3 \langle A \rangle^2/M$. Then the
mass matrices of the doublets and triplets in the ${\bf 10}_i$
are given by

\begin{equation}
M_2 = \left( \begin{array}{ccc}
0 & m & 0 \\m & 0 & m' \\ 0 & m' & 0 \end{array} \right), \;\;
M_3 = \left( \begin{array}{ccc}
0 & m & 0 \\m & 0 & m' \\ 0 & m' & a^2/M \end{array} \right).
\end{equation}

\noindent
Consequently, there is a
pair of ``massless" doublets, which are the MSSM Higgs, and two
supermassive pairs of doublets with determinant 
$\det' M_2 = m^2 + m^{'2}$. All three pairs of color-triplets
are supermassive, with determinant $\det M_3 = m^2 a^2/M$.
Thus, the contribution to $\epsilon_3$ is 
$\delta \epsilon_3 = (3 \tilde{\alpha}_G/5 \pi)$ 
$\ln \frac{m^2 a^2}{M_G M (m^2 + m^{'2})}$. To compute $F^{-1}$,
assume that at the GUT scale only the ${\bf 5}_1$ and 
$\overline{{\bf 5}}_1$ couple to the
quark and lepton multiplets, with Yukawa couplings $Y_1$ and
$\overline{Y}_1$. (Here and throughout it will often be convenient
to classify fields using $SU(5)$. We are not assuming, however,
that $SO(10)$ actually breaks to $SU(5)$.)
From the form of the doublet mass matrix in Eq. (7), it is
apparent that the light MSSM doublets are in the linear
combinations $(m' {\bf 5}_1 - m {\bf 5}_3)/\sqrt{m^2 + m^{'2}}$
and $(m' \overline{{\bf 5}}_1 - m \overline{{\bf 5}}_3)/
\sqrt{m^2 + m^{'2}}$. It is the VEVs of these combinations that
are the $v_2$ and $v_1$ of the MSSM and 
give mass to the light quarks and leptons.
Hence, the Yukawa couplings $Y_1$ and $\overline{Y}_1$, which control
proton decay, are proportional to $(m'/\sqrt{m^2 + m^{'2}})^{-1}$
times the light fermion mass matrices $M_U$ and $M_D$. 
Furthermore, the propagator of the
superheavy color-triplet Higgsinos is $(M_3^{-1})_{11} =
\frac{m^{'2}}{m^2 a^2/M}$. Consequently 
$F^{-1} = M_G \left( \frac{m^2 + m^{'2}}{m^{'2}} \right)
\left( \frac{m^{'2}}{m^2 a^2/M} \right) = \left( 
\frac{(m^2 + m^{'2}) M M_G}{m^2 a^2} \right)$. 
Comparing with the expression for $\epsilon_3$,
one sees that the doublet-triplet-splitting sector contributes here,
as it does for the simpler model, $\epsilon_3 = 
(3 \tilde{\alpha}_G/5 \pi) \ln F$.

This example shows how to generalize the result. Suppose that the doublets
and triplets, including the light doublets and the triplets that
contribute to proton decay, are contained in a set
of $\overline{{\bf 5}} + {\bf 5}$ Higgs fields labelled by 
the index $k$. Let the weak doublets in these multiplets be denoted
by $H_k$ and $\overline{H}_k$. Then the masses of the light
fermions are given by $M_U = \sum_k Y_k \langle H_k \rangle$ and
$M_D = M_L = \sum_k \overline{Y}_k \langle \overline{H}_k \rangle$,
where $Y_k$ and $\overline{Y}_k$ are Yukawa coupling matrices.
We assume for simplicity that the color-triplet Higgsinos have
the same Yukawa couplings as the doublet Higgs fields, as is 
the case in minimal SUSY $SU(5)$.

The mass matrices of the doublets and triplets in $\overline{{\bf 5}}_k
+ {\bf 5}_k$ will be denoted $M_2$ and $M_3$. Let $M_2$ be diagonalized
by unitary matrices $\overline{U}$ and $U$:
$\overline{U}^{\dag} M_2 U = \Lambda =$ diagonal. Unification of
gauge couplings requires that there be only one light pair of
doublets, $\overline{H}_0$ and $H_0$. The light eigenvalue of 
$\Lambda$ is therefore $\Lambda_0$, and the masses of the
light fermions are given by $M_U = \sum_k Y_k U_{k0} \langle H_0 
\rangle = (\sum_k Y_k U_{k0}) v_2$, and $M_D = \sum_k \overline{Y}_k 
\overline{U}_{k0} \langle \overline{H}_0 \rangle  = 
(\sum_k \overline{Y}_k \overline{U}_{k0}) v_1$. 

Consider the inverse of the matrix $M_2 = \overline{U} \Lambda U^{\dag}$.
The $kk'$ element of that inverse is given by
$(M_2^{-1})_{kk'} = (U \Lambda^{-1} \overline{U}^{\dag})_{kk'}
\cong U_{k0} \overline{U}_{k'0}^* \Lambda_0^{-1}$. The fact has been used
that $\Lambda_0^{-1}$ is much greater than all the other
$\Lambda_j^{-1}$ and hence dominates the sum. On the other hand,
$(M_2^{-1})_{kk'} = \det(m_2^{kk'})/\det M_2$, where $m_2^{kk'}$
is the cofactor of the $kk'$ element of $M_2$. Similarly,
$(M_3^{-1})_{kk'} = \det(m_3^{kk'})/\det M_3$, where $m_3^{kk'}$
is the cofactor of the $kk'$ element of $M_3$.
Thus,  

\begin{equation}
(M_3^{-1})_{kk'} = (M_2^{-1})_{kk'} \left( \frac{\det M_2}{\det M_3}
\right) \left( \frac{\det m_3^{kk'}}{\det m_2^{kk'}} \right).
\end{equation}

\noindent
Moreover, the Higgsino-mediated proton decay amplitude is proportional 
to the expression $\sum_{kk'} Y_k \overline{Y}_{k'}^*(M_3^{-1})_{kk'}$. 
Using the fact that $\det M_2 = \Lambda_0 \det' M_2$, as well as
the result just obtained that
$(M_2^{-1})_{kk'} \cong U_{k0} \overline{U}_{k'0}^* \Lambda_0^{-1}$, 
and Eq. (8), this expression can be written

\begin{equation}
\sum_{kk'} Y_k \overline{Y}_{k'}^* (M_3^{-1})_{kk'} =
\sum_{kk'} Y_k \overline{Y}_{k'} U_{k0} \overline{U}_{k'0}^*
\left( \frac{\det' M_2}{\det M_3} \right) 
\left( \frac{\det m_3^{kk'}}{\det m_2^{kk'}} \right).
\end{equation}

\noindent
Assume that $\left( \frac{\det m_3^{kk'}}{\det m_2^{kk'}} \right)
\equiv r_{kk'}$ is independent of $k$ and $k'$, and call it simply $r$.
Then the sum over $k$ and $k'$ reduces to 
$\sum_{kk'} Y_k \overline{Y}_{k'}^* U_{k0} 
\overline{U}_{k'0}^* = (M_U/v) (M_D/v')$, which, of course, is just
the same combination that 
appears in the proton decay amplitude in the minimal SUSY $SU(5)$ model.
Therefore, the suppression factor $F^{-1}$ is given by

\begin{equation}
F^{-1} = r M_G \left( \frac{\det' M_2}{\det M_3} \right).
\end{equation}

\noindent
This means that the contribution to $\epsilon_3$ of the doublets 
and triplets in the fields $\overline{{\bf 5}}_k
+ {\bf 5}_k$ is just

\begin{equation}
\delta \epsilon_3 |_{2/3} = \frac{3 \tilde{\alpha}_G}{5 \pi} \ln (r F).
\end{equation}

\noindent
Since $F$ must generally be large to sufficiently suppress proton
decay, it must be that $r$ ($\equiv \det m_3^{kk'}/\det m_2^{kk'}$)
is quite small. How can this happen? Some contributions to the
mass matrices will be $SU(5)$ invariant, and will therefore
contribute in the same way to the matrices $m_3^{kk'}$ and $m_2^{kk'}$.
If these are the only contributions, then $r = 1$. On the other
hand, a VEV which is non-singlet under the $SU(5)$ subgroup
of $SO(10)$, such as the VEV of an
adjoint, will generally contribute differently to
$m_3^{kk'}$ and $m_2^{kk'}$. Calling this VEV $\langle A \rangle$, one
can therefore write $m_3^{kk'} = m^{kk'}
+ \Delta_3^{kk'}(\langle A \rangle)$ and $m_2^{kk'} = m^{kk'}
+ \Delta_2^{kk'}(\langle A \rangle)$. 

If $r_{kk'}$ is to be
much less than one, as needed, one of two things obviously has to
be the case. (1) There must be a fortuitous relationship between the
magnitude of the elements of $m^{kk'}$ and those of 
$\Delta_3^{kk'}(\langle A \rangle)$ so that the determinant of
the sum of those two matrices nearly vanishes. This requires
a fine-tuning. Or, (2) there must be enough entries in $m_3^{kk'}$ 
which get vanishing
contributions from {\it both} $m^{kk'}$ and $\Delta_3^{kk'}
(\langle A \rangle)$ that $m_3^{kk'}$ has vanishing determinant
without fine-tuning. However, in this case, $m_2^{kk'}$ will {\it also} 
have vanishing determinant unless certain elements of 
$\Delta_2^{kk'}(\langle A \rangle)$ are
non-zero even though the corresponding elements of 
$\Delta_3^{kk'}(\langle A \rangle)$ vanish. 

The foregoing argument implies that for proton decay to be greatly 
suppressed without fine-tuning and without giving a contribution to 
$\epsilon_3$ that contains a
large positive logarithm, {\it there must be a superheavy VEV
which can contribute to doublet masses while not 
contributing to the mass of their color-triplet partners}. 
This VEV is presumably that of an adjoint.

It is easily
shown that this cannot happen if the only superlarge VEV which breaks 
the $SU(5)$ subgroup of $SO(10)$ is an adjoint pointing in the $B-L$
direction. However, it can happen if there are two (or more) 
adjoints, with one of them having a VEV
proportional to $B-L$ and the other a VEV proportional to $I_{3R}$
(that is, the third generator of the $SU(2)_R$ subgroup of $SO(10)$). 
In fact, such a situation
was proposed in [7], where the following type of structure was suggested
as a possibility: $W_{2/3} = g T_1 A T_2 + T_2 B^2 T_2/M$. (Compare
with Eq. (2).) Here $A$ and $B$ are adjoints and $\langle A \rangle
\propto B-L$ and $\langle B \rangle \propto I_{3R}$. Then the doublet
and triplet mass matrices are of the form

\begin{equation}
M_2 = \left( \begin{array}{cc}
0 & 0 \\ 0 & b^2/M \end{array} \right), \;\;
M_3 = \left( \begin{array}{cc}
0 & a \\ -a & 0 \end{array} \right).
\end{equation}

\noindent
In this case, if only $T_1$ couples to the quarks and leptons,
$F^{-1} = M_G (M_3^{-1})_{11} = 0$, while the logarithm in
$\delta \epsilon_3$ is $\ln (a^2 M/b^2 M_G)$, which can take any value,
and is naturally of order one. The reason this structure achieves
the desired result is that the cofactor $m_3^{11} = (M_3)_{22} = 0$, 
while the cofactor
$m_2^{11} = (M_2)_{22} = b^2/M$. That is, as our previous
reasoning indicated was needed, the VEV of $B$ contributes to
the 22 entry of the doublet matrix without contributing to the
22 element of the triplet matrix.

This shows that one can suppress proton decay without making the 
problem of $\alpha_s$ worse than it is in minimal SUSY $SU(5)$. 
Moreover, it is also possible to get the
contribution of $\delta \epsilon_3 \approx -0.02$ which is needed even
in minimal SUSY $SU(5)$ by assuming the ratio $a^2 M/b^2 M_G$ to be
about $10^{-1}$ to $3 \times 10^{-2}$. This is the idea
suggested in [12].

\section{Solution to the problem of $\alpha_s$ with one adjoint}

If the $SU(5)$ subgroup of $SO(10)$ is broken at unification
scale only by a single adjoint, then it is not as straightforward
to solve the problem of $\alpha_s$ with split multiplets.
Nevertheless, there are solutions, both in the case that the sole
adjoint points in the $B-L$ direction, and in the case that it
points in the $I_{3R}$ direction. 
\vspace{0.2cm} 

\noindent
(I) $\langle A \rangle \propto B-L$:

An adjoint VEV that points in the $B-L$ direction can give mass to
triplets that are in ${\bf 10}$'s of $SO(10)$, without giving
mass to doublets. This the idea behind the Dimopoulos-Wilczek
(or missing VEV) mechanism. But there are no representations in
which it does the opposite, giving mass to doublets but not triplets.
This means that the suppression of proton decay in this case will
necessarily lead to a significant positive contribution 
to $\epsilon_3$. However, it is possible with
$\langle A \rangle \propto B-L$ for there to be other kinds
of split multiplets which compensate by giving large negative
contributions to $\epsilon_3$.

The simplest possibility is an extension of the mechanism
proposed in [13]. In that paper the following structure was
proposed for coupling pairs of spinors (${\bf 16} +
\overline{{\bf 16}}$) to an adjoint:

\begin{equation}
W_{CA} = \overline{C}(PA/M_1 + Z_1)C' + \overline{C}'(P A/M_2 + Z_2) C.
\end{equation}

\noindent
Here $C + \overline{C}$ are a conjugate pair of spinors which get 
superlarge VEVs in the $SU(5)$-singlet direction, and $C' +
\overline{C}'$ are a conjugate pair of spinors that get no
superlarge VEVs. The fields $P$, $Z_1$ and $Z_2$ are gauge
singlets which have superlarge expectation values. It is assumed
that the quantum numbers of the $Z_i$ are such that they have no
other couplings. (This is the purpose of the presence of the field
$P$. Were it not there, $Z_i$ and $A$ would have the same non-$SO(10)$
quantum numbers, and other couplings of the $Z_i$'s would be possible.) 
The significance of this structure, as explained in [13], 
is that it allows the coupling of the adjoint $A$ to the spinors 
$C + \overline{C}$, which is required to avoid disastrous colored 
pseudo-goldstone bosons, without destabilizing the VEV of the adjoint. 
In particular, these terms do not destabilize the required form of 
$\langle A \rangle$
by contributing to $F_A$, since their contribution to $F_A$ vanishes
due to the vanishing of the VEVs of $C'$ and $\overline{C}'$. The
$Z_i$ are effectively ``sliding singlets", which adjust to make 
$F_{C'}$ and $F_{\overline{C}'}$ vanish, and thus avoid breaking
supersymmetry at the GUT scale.

Consider the equation $F_{\overline{C'}}^* = 0$. This implies that
$(PA/M_2 + Z_2) C = 0$. Since $\langle A \rangle = \frac{3}{2} (B-L)$
(see Eq. (1)), it follows that $\langle Z_2 \rangle = - \frac{3}{2} a
(B-L)_{\langle C \rangle} P/M_2$. (Here we mean by $B-L$ the $SO(10)$
generator that acts on the quarks and leptons in a spinor like $B-L$.) 
The component of $C$ that gets a VEV has the same 
$SO(10)$ quantum numbers as a left-handed antineutrino, and hence
$(B-L)_{\langle C \rangle}
= -1$. Thus $(PA/M_2 + Z_2) = \frac{3}{2} a (B-L + 1) P/M_2$.
What is interesting about this for our present purposes is that
$(B-L +1)$ vanishes not only for left-handed antineutrinos
but also for left-handed ``positrons", that is, for the components
of a spinor which are in $({\bf 1}, {\bf 1}, +1)$'s of $SU(3) \times 
SU(2) \times U(1)$. As we shall see, this can lead to massless
(or at least very light) fields which have those quantum numbers.
The virtue of this is that such a field makes a sizable negative
contribution to $\epsilon_3$.

For example, suppose the structure in Eq. (13) is extended in the 
following way:

\begin{equation}
W_{CA} = \sum_{i=1}^2 \overline{C}_i (PA/M_i + Z_i)C'_i + 
\sum_{i = 1}^2 \overline{C}'_i (P A/M'_i + Z'_i) C_i,
\end{equation}

\noindent
where dimensionless couplings have been suppressed. Here there are four
spinor-antispinor pairs, instead of two. It is easily checked that
in addition to the goldstone modes that are eaten by the Higgs
mechanism, there is one pair of goldstones with the $SU(3) \times
SU(2) \times U(1)$ quantum numbers $({\bf 1}, {\bf 1}, \pm 1)$. These
goldstones can get mass from higher-dimensional operators. The
mass they get is therefore highly model dependent. Let us call it $m_1$.
The the contribution of these pseudo-goldstones to $\epsilon_3$
is at one loop $\delta \epsilon_3 = - \frac{3 \tilde{\alpha}_G}{10 \pi} 
\ln (M_G/m_1)$. If $m_1 \sim 10^{11}$ GeV, then this contributes
$\delta \epsilon_3 \cong -0.04$, while if $m_1 \sim 10^8$ GeV, 
$\delta \epsilon_3 \cong -0.08$. 

It is possible that these pseudo-goldstone particles can even be as
light as the weak scale. This would be an interesting signal of
unification physics. Even if these pseudos are at some intermediate
scale they may play a cosmological role.

\vspace{0.2cm}

\noindent
(II) $\langle A \rangle \propto I_{3R}$

Suppose that there is only a single adjoint and its VEV points in the
$I_{3R}$ direction. That is,

\begin{equation}
\langle A \rangle = \left( \begin{array}{ccccc}
a & & & & \\ & a & & & \\ & & 0 & & \\ & & & 0 & \\
& & & & 0 \end{array} \right) \times i \tau_2.
\end{equation}

\noindent
In this case the Dimopoulos-Wilczek or missing-VEV
mechanism cannot work as in Eq. (2), for an adjoint with a VEV
proportional to $I_{3R}$ will give mass to the weak doublets in 
a ${\bf 10}$ of $SO(10)$ while leaving the triplets massless.
This is the just the reverse of what is needed to solve the 
doublet-triplet splitting problem. However, in [14] it was pointed out
that an adjoint VEV proportional to $I_{3R}$ can leave massless 
the weak doublets that are in {\it spinors} of $SO(10)$, while making their
color-triple partners superheavy. The reason for this is that those 
weak doublets have $I_{3R} = 0$.
This is obvious, since such doublets have the same gauge charges as 
a lepton doublet, and hence are singlets under $SU(2)_R$. 

The models suggested in [14] actually have more than one adjoint, 
but it is a simple matter to modify those models so that only a single
adjoint is needed. Consider the following superpotential:

\begin{equation}
\begin{array}{ccl}
W_{2/3} & = & m_1 \overline{{\bf 16}}_1 {\bf 16}_1 + 
m_2 \overline{{\bf 16}}_2 {\bf 16}_2 \\
& & {\bf 16}_1 {\bf 10} \langle {\bf 16} \rangle 
+ \overline{{\bf 16}}_2 {\bf 10} \langle \overline{{\bf 16}} \rangle
+ \overline{{\bf 16}}_1 {\bf 16}_2 \langle {\bf 45} \rangle. 
\end{array}
\end{equation}

\noindent
The sector that contains the light Higgs doublets consists of
a vector ${\bf 10}$ and two conjugate pairs of spinors
$\overline{{\bf 16}}_1 + {\bf 16}_1 + \overline{{\bf 16}}_2
+ {\bf 16}_2$. Denote the superlarge VEVs of the 
$SU(5)$-singlet components of ${\bf 16}$ and $\overline{{\bf 16}}$, 
by $\Omega$ and $\overline{\Omega}$ respectively. Then the mass matrix
of the $SU(5)$ $\overline{{\bf 5}}$'s has the following form:

\begin{equation}
(\overline{{\bf 5}}_{16_1}, \overline{{\bf 5}}_{16_2}, 
\overline{{\bf 5}}_{10}) \left( \begin{array}{ccc}
m_1 & 0 & \Omega \\ \langle A \rangle & m_2 & 0 \\
0 & \overline{\Omega} & 0 \end{array} \right) \left(
\begin{array}{c} {\bf 5}_{\overline{16}_1} \\
{\bf 5}_{\overline{16}_2} \\ {\bf 5}_{10} \end{array} \right).
\end{equation}

\noindent
In the doublet mass matrix, $M_2$, the entry $\langle A \rangle$
vanishes, since, as we noted, the doublets in the spinors have
$I_{3R} = 0$. Thus $M_2$ has a zero eigenvalue. Projecting this
out, $\det' M_2 = (m_1^2 + \Omega^2)^{1/2} (m_2^2 +
\overline{\Omega}^2)^{1/2}$. On the other hand, all the eigenvalues
of the mass matrix of the triplets are superheavy, and
$\det M_3 = a \Omega \overline{\Omega}$. Thus the desired 
doublet-triplet splitting is achieved. Note also that the $SU(5)$
${\bf 10}$'s in the spinors are all made superheavy by the VEVs
$\Omega$ and $\overline{\Omega}$. 

Suppose, as is simplest, that only the ${\bf 10}$ of Higgs fields
couples to the quarks and leptons. Then it is easy to see that the
conditions of the theorem proved in section 3 are satisfied, since 
only one $r_{kk'}$ is relevant. Thus $\delta \epsilon_3 = 
\frac{3 \tilde{\alpha}_G}{5 \pi} \ln F$, where $F^{-1}$ is the
factor by which the proton-decay amplitude is suppressed relative to
minimal $SU(5)$. Thus, so far, we have only seen how doublet-triplet
splitting may arise, but have not solved the problem of $\alpha_s$.
Now, however, it is possible to have ``upside-down" split multiplets that
give the needed large negative contributions to $\epsilon_3$.
These can arise from the coupling of the adjoint to {\it other} vectors
of Higgs fields, which we will denote $T'$ and 
$T^{\prime \prime}$. If there are terms $T' A  T^{\prime \prime} 
+ m' T^{\prime 2} + m^{\prime \prime} T^{\prime \prime 2}$, where
$m' m^{\prime \prime} \ll a^2$, for example, then in these multiplets
the doublets are much heavier than the triplets and they contribute
$\delta \epsilon_3 \cong  -\frac{3 \tilde{\alpha}_G}{ 5 \pi} 
\ln( a^2/m'm^{\prime \prime})$. This can easily be made large
enough to give a satisfactory agrement for $\alpha_s$.

One can see that in both the cases with one adjoint Higgs the
sector which gives doublet-triplet splitting substantially
increases $\alpha_s$, and that agreement with experiment is
achieved by assuming that another sector decreases $\alpha_s$
by a roughly equal amount. In that case, the fact that the
gauge couplings appear to unify so accurately in the MSSM seems
to be the result of a fortuitous cancellation. On the other
hand, in the two-adjoint case considered at the end of section 3
the doublet-triplet sector does not tend to give large 
contributions to $\alpha_s$, and therefore the agreement with
experiment can happen in a manner that appears less contrived.

\section{Conclusion}

We have shown that in a very general class of supersymmetric
unified models based on $SO(10)$ the sector that produces the
doublet-triplet splitting contributes a positive amount to
$\epsilon_3$ and thus worsens the discrepancy in $\alpha_s$
between the simplest models and experiment.
We have also shown that with two adjoints, whose VEVs are
proportional to $B-L$ and $I_{3R}$, it is easy to arrange that 
the doublet-triplet-splitting sector does not exacerbate the
discrepancy in $\alpha_s$, and even cures it by having 
``upside-down" split multiplets (that is,
split multiplets where the doublets are much heavier than the 
triplets) that contribute negatively to $\epsilon_3$ and thus
bring $\alpha_s$ back into agreement. We have seen why
the problem of $\alpha_s$ is not as straightforward to solve 
if there is only a single adjoint. Nevertheless, we have found 
that a technically natural solution can be found both in the case where
the VEV of the single adjoint points in the $B-L$ direction,
and the case where it points in the $I_{3R}$ direction.
In these cases, however, the doublet-triplet-splitting sector
makes the problem worse, and another sector must happen to
contribute in a way that compensates for it.

One of the interesting features of the solution with a single
adjoint whose VEV is proprtional to $B-L$ is the appearance
of pseudo-goldstone fields with the standard model quantum numbers 
$({\bf 1}, {\bf 1}, \pm 1)$. These can conceivably be at the weak 
scale. The possibility of a single adjoint whose VEV is proportional
to $I_{3R}$ deserves further study. Important questions for this case
are whether the gauge hierarchy is easily stabilized against the effects
of higher-dimensional operators, and whether realistic Yukawa
structures for the quarks and leptons are possible.

\section*{References}

\begin{enumerate}

\item S. Dimopoulos, S. Raby, and F. Wilczek, {\it Phys. Rev.}
{\bf D24}, 1681 (1981); L. Ibanez and G. Ross, {\it Phys. Lett.}
{\bf 105B}, 439 (1981); U. Amaldi, A. Bohm, L. Durkin, P. Langancker,
A. Mann, W. Marciano, A. Sirlin, and H. Williams, {\it Phys. Rev.}
{\bf D36}, 1385 (1987); P. Langacker and M. Luo, {\it Phys. Rev.}
{\bf D44}, 817 (1991). 
\item P. Langacker and N. Polonsky, {\it Phys. Rev.}{\bf D47}, 4028
(1993).
\item J. Bagger, K. Matchev, and D. Pierce, {\it Phys. Lett.} 
{\bf 348B}, 443 (1995).
\item B.R. Webber, in {\it Proceedings 27th International Conference
on High Energy Physics}, Glasgow, Scotland, 1994, edited by P.J. Bussey
and I.G. Knowles (IOP, London, 1995).
\item V. Lucas and S. Raby, {\it Phys. Rev.} {\bf D54}, 2261 (1996).
\item R. Arnowitt and P. Nath, {\it Phys. Rev. Lett.} {\bf 69},
725 (1992); J. Hisano, H. Murayama, and T. Yanagida, {\it Phys. Rev. Lett.} 
{\bf 69}, 1014 (1992); {\it Nucl. Phys.} {\bf B402}, 46 (1993);
K. Hagiwara and Y. Yamada, {\it Phys. Rev. Lett.} {\bf 70}, 709 (1993);
Y. Yamada, {\it Z. Phys.} {\bf C60}, 83 (1993).
\item K.S. Babu and S.M. Barr, {\it Phys. Rev.} {\bf D48}, 5354 (1993).
\item B. Brahmachari and R. Mohapatra, {\it Int. J. Mod. Phys.}
{bf A11}, 1699 (1996) contains a good review of the problem and
of the various solutions that have been proposed.
\item L. Roszkowski and M. Shifman, {\it Phys. Rev.} {\bf D53}, 404 (1996);
S. Raby, hep-ph/9712254.
\item D.G. Lee and R.N. Mohapatra, {\it Phys. Rev.} {\bf D52}, 4125 (1995).
B. Brahmachari and R.N. Mohapatra, {\it Phys. Lett.} {\bf 357B}, 566 (1995);
S.P. Martin and P. Ramond, {\it Phys. Rev.} {\bf D51}, 6515 (1995).
\item S. Dimopoulos and F. Wilczek, in {\it The Unity of the
Fundamental Interactions}, Proceedings of the $19^{{\rm th}}$ Course
of the International School of Subnuclear Physics, Erice, Italy, 1981,
edited by A. Zichichi (Plenum, New York, 1983).
\item M. Bastero-Gil and B. Brahmachari, {\it Phys. Rev.} {\bf D54},
1063 (1996).
\item S.M. Barr and S. Raby, {\it Phys. Rev. Lett} {\bf 79}, 4748 (1997).
\item G. Dvali and S. Pokorski, {\it Phys. Lett.} {\bf 379B}, 126 (1996).

\end{enumerate}

\end{document}